# FlowMapper.org: A web-based framework for designing origin-destination flow maps

Caglar Koylu[1]*, Geng Tian[1], Mary Windsor

*Corresponding author: caglar-koylu@uiowa.edu

1 Geographical and Sustainability Sciences, University of Iowa, Jessup Hall, Iowa City, IA 52242

## Abstract

FlowMapper.org is a web-based framework for automated production and design of origin-destination flow maps (https://flowmapper.org). FlowMapper has four major features that contribute to the advancement of existing flow mapping systems. First, users can upload and process their own data to design and share customized flow maps. The ability to save data, cartographic design and map elements in a project file allows users to easily share their data and/or cartographic design with others. Second, users can customize the flow line symbology by including options to change the flow line style, width, and coloring. FlowMapper includes algorithms for drawing curved line styles with varying thickness along a flow line, which reduces the visual cluttering and overlapping by tapering flow lines at origin and destination points. The ability to customize flow symbology supports different flow map reading tasks such as comparing flow magnitudes and directions and identifying flow and location clusters that are strongly connected with each other. Third, FlowMapper supports supplementary layers such as node symbol, choropleth, and base maps to contextualize flow patterns with location references and characteristics such as net-flow, gross flow, net-flow ratio, or a locational attribute such as population density. FlowMapper also supports user interactions to zoom, filter, and obtain details-on-demand functions to support visual information seeking about nodes, flows and regions. Finally, the web-based architecture of FlowMapper supports server-side computational capabilities to process, normalize and summarize large flow data to reveal natural patterns of flows.

# 1. Introduction

Flow maps illustrate movements of tangible and intangible phenomenon between locations. While there are many types of flow maps, origin-destination (OD) flow maps illustrate directed flows between origin and destination locations while ignoring the actual routes between locations (Slocum et al., Dent, Torguson, & Hodler, 2008; 2009; Tobler, 1987). There have been recent advancements in web-based flow mapping applications. Stephen & Jenny (2017) designed a flow map application that allows users to obtain an overview of flow patterns in U.S. domestic migration. State-to-state migration flows are visualized based using Jenny *et al.'s* (2017) force-directed layout. The application supports details-on-demand functionality by allowing users to select and visualize county-level flows between the selected state and other states using a circular layout. Jenny *et al.'s* (2017) force-directed layout that is guided by cartographic principles for flow mapping is also avaiable in Java (Jenny, 2019). Nost et al. (2017) developed HazMatMapper to explore potential environmental justice issues by visualizing hazardous waste flow data. Using HazMatMapper, users can acquire an overview first then filter the dataset in greater detail with a regional view and obtain details-on-demand with an information panel that provides additional statistics. Flowmap.blue is a web-based flow mapping application powered by WebGL (Parisi, 2012) that allows rendering, interactive filtering and animation of a large number of flow lines and nodes using GPU computation (Ilya Boyandin, 2021). Its interface provides a variety of base maps and straight flow lines with half-arrows at the end points to depict flow direction. However, curved flow lines are not available in Flowmap.blue. Flowmap.blue includes a location clustering feature to aggregate flows between clusters that are formed by nearby units. Location clustering reduces visual cluttering of flows and allows visualization of flow patterns at different scales (e.g., zoom levels). However, the clustering method suffer from the modifiable areal unit problem that group nearby locations while disregarding flow connections (Zhu et al., 2019). Kepler.gl is another WebGL-powered platform that can render millions of points and thousands of arc lines to produce flow maps for large data (He, 2018). Kepler.gl can perform spatial aggregations on the fly, and visualize spatial features such as points, paths, polygons, grids and hexbins, as well. Both Flowmap.blue and Kepler.gl include temporal filtering to generate animations for temporal flow data and geographic filtering to highlight flow lines (arc) from and to selected origin and destination locations. Despite its scalability to visualize large number of features, Kepler.gl offers a limited number of symbology and scaling choices. Also, flow lines are monotone curved arcs, and there is little control over basic cartographic elements and principles such as flow line style, contextual layers, base maps, and references, map elements, etc. Despite these new developments there is still much work to be done to develop open-source flow mapping systems that can process and normalize flow data with computational capabilities, and design and share customized flow maps in professional cartographic workflow with greater customizability and design flexibility.

In this article, we introduce a web-based framework, https://flowmapper.org, to create and share OD flow maps in an interactive environment. Our goal is to provide users greater customizability and design flexibility that could help produce an interactive visualization and a professional static map product that requires minimal touch-up in other graphics software. FlowMapper has four major contributions to web-based flow mapping. First, FlowMapper allows users to upload and process their own data to produce, customize and share flow maps. The ability to store data, cartographic design and map elements in a project file allows users to easily share their data and/or styling with others. Second, users can customize the flow line symbology by including options to change the flow line style, width, and coloring. FlowMapper includes algorithms for drawing curved line styles with varying thickness along a flow line, which reduces the visual cluttering and overlapping by tapering flow lines at origin and destination points. The ability to customize flow symbology supports different flow map reading tasks such as comparing flow magnitudes and directions and identifying flow and location clusters that are strongly connected with each other. Third, FlowMapper supports supplementary layers such as node symbol, choropleth, and base maps to contextualize flow patterns with location references and characteristics such as net-flow, gross flow, net-flow ratio, or a locational attribute such as population density and median income. FlowMapper supports user interactions to zoom, filter, and obtain details-on-demand functions to support visual information seeking about nodes, flows and regions. Finally, the web-based architecture of FlowMapper supports server-side computational capabilities to process, normalize and summarize large flow data to reveal natural patterns of flows.

## 2. Related Work

There are several remaining challenges for developing effective flow mapping applications. First, flow map comprehension is challenging due to the visual cluttering of flows, even in small data sets. For example, interstate migration within the U.S. between 2015 and 2020 form a network of state-to-state migration flows, which consists of 50 nodes (states), thousands of links and millions of migrants moving through those links between the states. Displaying all flows makes it impossible for the map reader to reveal flow patterns. Filtering is used to address visual cluttering by displaying a subset of flows that exceed a threshold value, or the top rank flows, or flows to and from a selected location (Tobler, 1987). However, filtering lacks the ability to provide an overview of flow patterns. Edge bundling is commonly applied to bundle flow lines to reduce cluttering and improve clarity (Graser et al., 2019; D. Holten & van Wijk, 2009; Phan et al., 2005). However, bundled edges may be perceived as actual routes of flows, which also makes it difficult for the map reader to follow the connections between origins and destinations (von Landesberger et al., 2016). To avoid visual cluttering of flows, OD data can be visualized as a matrix where rows represent the locations of flow origins and columns the locations of destinations (Ghoniem, Fekete, & Castagliola, 2005). Wood *et al.* (2010) introduced OD map based on a matrix visualization that preserves the spatial layout of all origin and destination locations and reveal spatial associations between pairs of locations. Yang *et al.* (2017) expanded the OD map (Wood et al., 2010) by integrating geographic embedding and several types of interaction including highlighting, filtering and zooming. However, complex connections between locations are hard to identify in matrix and the OD map visualizations.

Flow maps convey users different types of information such as flow magnitude, directions, clustering, and spatial focusing. These different types of information often require different symbology choices to best communicate the type of pattern to users. It is difficult to design complex interfaces that can guide users in choosing a symbology for a particular task (Koylu & Guo, 2017). Moreover, designing a user-friendly and accessible interface for the general public is challenging due to the interface complexity generated by enabling various design alternatives (Vincent et al., 2019). As compared to straight flow lines, curved flow lines have been shown to enhance readability in graph visualization (Holten et al., 2010) and flow maps (Jenny et al., 2018; Koylu, 2014). A flow map eye tracking study (Dong et al., 2018) revealed that users could more accurately interpret curved flows over straight flows. Dong *et al.* (2018) also found that the use of a color gradient to represent flow volume were more effective for user's interpretation of flows that differed by line thickness. Jenny *et al.* (2018) outlined design principles for curving of flows in origin-destination flow maps. Major principles include minimizing flow line intersections, obtaining large angles at intersections, drawing symmetric single and gradual curves, avoiding flows to pass through nodes and narrow angles between flows at shared nodes. Jenny *et al.* *(2017)* used a force-directed graph drawing algorithm to follow these principles and create OD flow maps that depict a one-way connection (net flow) between two locations. The force-directed algorithm used by Jenny *et al.* (2017) is less effective for reducing visual cluttering in two-way directed flow maps in which there are two flow connections between a pair of locations. This is because the graph drawing algorithm cannot effectively support the design principles when there are too many flow lines and especially two flow lines between the same pair of nodes. In addition, although symmetric curves may enhance flow map readibility for identifying flow connections, magnitudes, and clustering, other user study findings suggest that asymmetric curvature enhances the perception of flow directions (Koylu & Guo, 2017; Ware, Kelley, & Pilar, 2014). Therefore, the best use of flow symbolization may depend on the data and scale because both affect how a map reader scans the map and which features become visually salient.

Computational methods such as flow-based regionalization (Guo, 2009), location-based clustering (Andrienko et al., 2017; von Landesberger et al., 2016), and flow data smoothing and clustering (Guo & Zhu, 2014; Tao & Thill, 2016; Zhu et al., 2019) are used to summarize flows and reveal hidden patterns in flow data. Clustering and regionalization methods aggregate individual flows into flows between contiguous regions or clusters of locations and reduce the number of regions and flows to be displayed. Flow clustering and density estimation methods also reduce the number of flows by identifying and visualizing representative or significant flows rather than displaying all flows. Flow clustering methods show great promise for summarizing large flow data sets, providing an overview of patterns with multi-scale mapping ability. However, these methods are computationally expensive, and may be provided as asynchronous processing tools on a web-based flow mapping environment. In addition to summarizing flows in geographic space, researchers developed new methods to extract temporal flow trends (Andrienko et al., 2017; Boyandin et al., 2011; von Landesberger et al., 2016). Main challenge in time-variant flow mapping is the large number of spatial units and the difficulty of detecting changes between time periods. Boyandin *et al.* (2011) implemented a heat map to visualize the changes in flow magnitudes or node measures such as inflows and outflows between two time periods. von Landesberger et al. (2016) used the similarity of flow structures to cluster and decrease the number of time periods to be compared. Similarly, Andrienko et al. (2017) employed a temporal abstraction technique that clusters time periods based on the similarity of flows. Their method also allows clustering of flows with a common origin or a common destination by direction and distance ranges of flows using glyphs. However, it is difficult to perceive the connections between origins and destination pairs by comparing glyphs.

## 3. Map design

To support a variety of flow map reading tasks, FlowMapper consists of a flow map with alternative flow line designs, a node (point) symbol map, a region (choropleth) map, and a reference base map. Our goal is to help users design and share their customized flow maps but also export their final polished maps with minimal touch-up in other graphics software.

## 3.1.    Flow map

FlowMapper allows users to customize the flow line symbology by including options to change the flow line style, width (thickness), and coloring.

### 3.1.1.    Flow line style

We introduce algorithms for drawing straight and curved flow paths. Unlike monotone line thickness along the flow line, we reduce the overlapping of flow lines at the origin and destination locations by tapering the flow lines to thin points at the extremities. FlowMapper currently provides four flow line designs: Curve with half-arrow, tapered curve, teardrop curve, and straight-line with half-arrow (Figure 1). Curved paths are asymmetric to enhance readability of directions in two-way flow maps (Koylu & Guo, 2017). All flow paths are connected to the outer edge of the circles at each flow's start and end points instead of the center of circles. The algorithms and directions for producing the cubic Bezier flow symbols are explained in detail in Appendices Section 1.

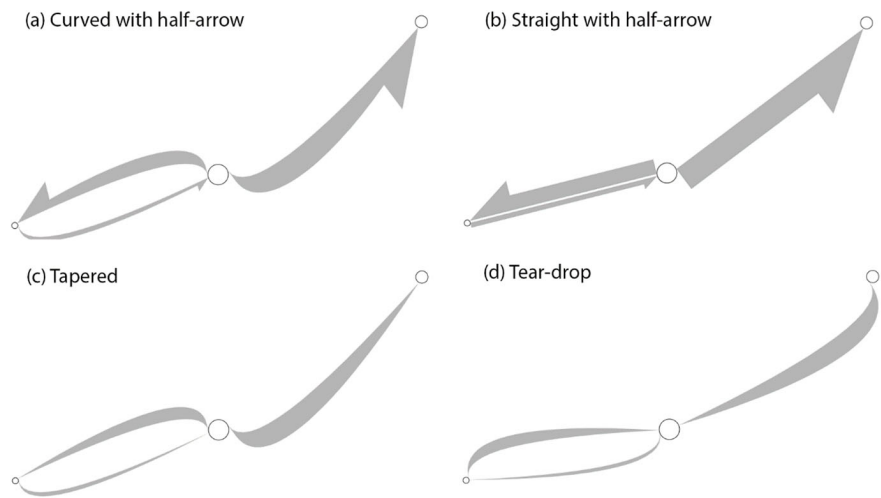

**Figure 1: Flow line styles (a) Curved path with half-arrow (b) Straight path with half-arrow (c) Tapered path (d) Teardrop path**

### 3.1.2.    Flow line (path) thickness

Users can choose a proportional scale to make flow thickness proportional to flow magnitude. Perceptual issues related to proportional symbol mapping are also inherent in proportional flow thickness - it is difficult to perceive differences in thickness between proportionally scaled flows. Moreover, longer flows are visually more salient than shorter flows, which is a similar problem to the perception of large areas in choropleth maps. Also, there are only a few large flows and many small flows in most flow data sets, which makes it challenging for comparing flow magnitudes. Similar to graduated (or range-graded) point symbols, we provide classification methods such as natural breaks, equal interval, quantile, and manual classification for classifying flow magnitudes.

### 3.1.3.    Flow line color

Users can select a single color or a color gradient to fill the area under the curve as a double visual variable (in addition to flow line width) to visualize flow magnitude. For the color gradient, users can either select a classification method with a color scheme adopted from colorbrewer.org or use an unclassed (or continuous) color scheme using min-max scaling of flow magnitudes between two selected colors. Users may use a flow stroke, which may enhance the perception of flow lines on areas where multiple flow lines cross over. Figure 2 illustrates proportional symbol legends with a single color (Figure 2.a) and a continuous (unclassed) color scheme from white to blue with a gray flow stroke (Figure 2.b). Proportional symbol legends include three values: minimum, average, and maximum. Figure 2.c illustrates a classified flow thickness and color with five classes using a quantile classification.

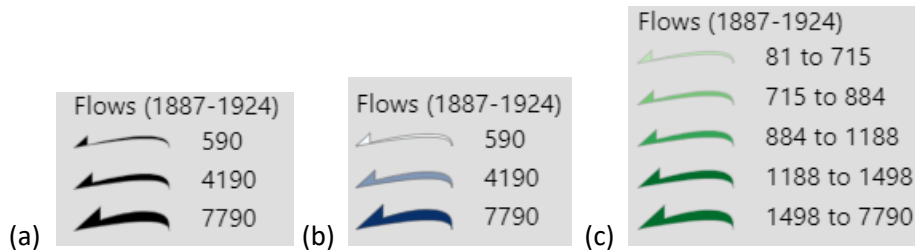

**Figure 2: Scaling and legend alternatives for flow symbols (a) a single color (black) flow symbol with line thickness scaled proportional to flow volume. (b) color and line thickness scaled proportional to flow volume (c) a classified flow thickness and color with five classes using a quantile classification.**

### 3.2. Node symbol map

Nodes are the coordinates of origin and destination locations for flows. While a node data set with id and coordinates is required to produce a flow map, users may choose to hide node symbols by using zero stroke-width and full transparency for fill color. The choice of whether to use node symbols depend on the data set and task. For example, node symbols are good anchors for airport locations for visualizing passenger flows from and to airports. On the other hand, hiding node symbols for visualizing flows between regions such as states may be preferred to highlight the nature of region-to-region flows.

### 3.3. Region map

A region choropleth map is an optional map for contextualizing of flow patterns with regional flow or attribute characteristics. Region maps are especially important when flow data is from regions-to-regions such as state-to-state migration flows. It is beneficial to use a region map to illustrate a location-based measure of flows such as net-flow ratio, migration efficiency, or a locational attribute such as population density and income.

### 3.4. Base map, map projections, and map elements

FlowMapper supports base maps produced by ESRI, Stamen and OpenStreetMap. For ESRI base maps, users may choose to add reference labels for places. Following projections are available in FlowMapper: Albers Equal Area projection is available for U.S., Africa, Australia, China, Europe, and South America, Robinson, Gall-Peters, and Mercator. Map elements such as map title, north arrow, projection label, and custom place labels are also available under base map settings (see Appendices Section 3).

### 3.5. Computational tools

FlowMapper currently supports two functions for processing flow and location data under Tools menu: "Polygons to Points" and "Normalize Flows". "Polygons to Points" tool transforms a polygon JSON file to a (node) point csv file with latitude and longitude coordinates of the geometric centroid of each polygon. "Normalize Flows" tool takes an input flow csv file and transforms it to a modularity flow file using a null expectation model. Flow normalization helps remove the effect of size differences and the ability for nodes to generate flows by calculating the difference between the actual flow and the expected volume of flow for each pair of locations (nodes). We explain the process for transforming raw flows to modularity flows in the Appendices Section 2.

## 4. Map use cases

We demonstrate the utilities of Flow Mapper using three scenarios that portray distinct flow map data sets, map extents, scales, and layouts.

### 4.1.  Scenario 1: Banana trade within South America in 2019

Figure 3 illustrates banana trade flows between countries in South America from 2019 (FAO, 2021). The choropleth map illustrates the net banana trade ratio, which is calculated by:

$$Net\ trade\ ratio\ (i) = 100{,}000 * \big(import(i) + export(i)\big)/population(i)$$

For each country i, we divide the sum of its imports and exports in tons by its population and multiply the division with 100,000 to derive the net banana trade ratio in tons per 100 thousand population. The countries with a negative net trade ratio are depicted in purple hue, which indicates larger export of bananas then imports in tons. On the other hand, the countries with positive net trade ratio are illustrated in green, which import more bananas than they export. We visualize the gross volume of banana trade in tons (import + export) using a graduated node (point) symbol map with three classes, which are placed at approximately the centroid of country boundaries. Finally, we visualize the banana trade from a country to another country using graduated flow symbols with four classes. Countries with a warmer climate that are closer to the equator export more bananas to countries in the south of the continent that have less suitable climate for producing bananas. From the nodes showing the gross volume of banana trade, we can see that Ecuador exports more bananas to other countries than Brazil and Peru while countries like Argentina and Chile are the major importers of bananas within South America.

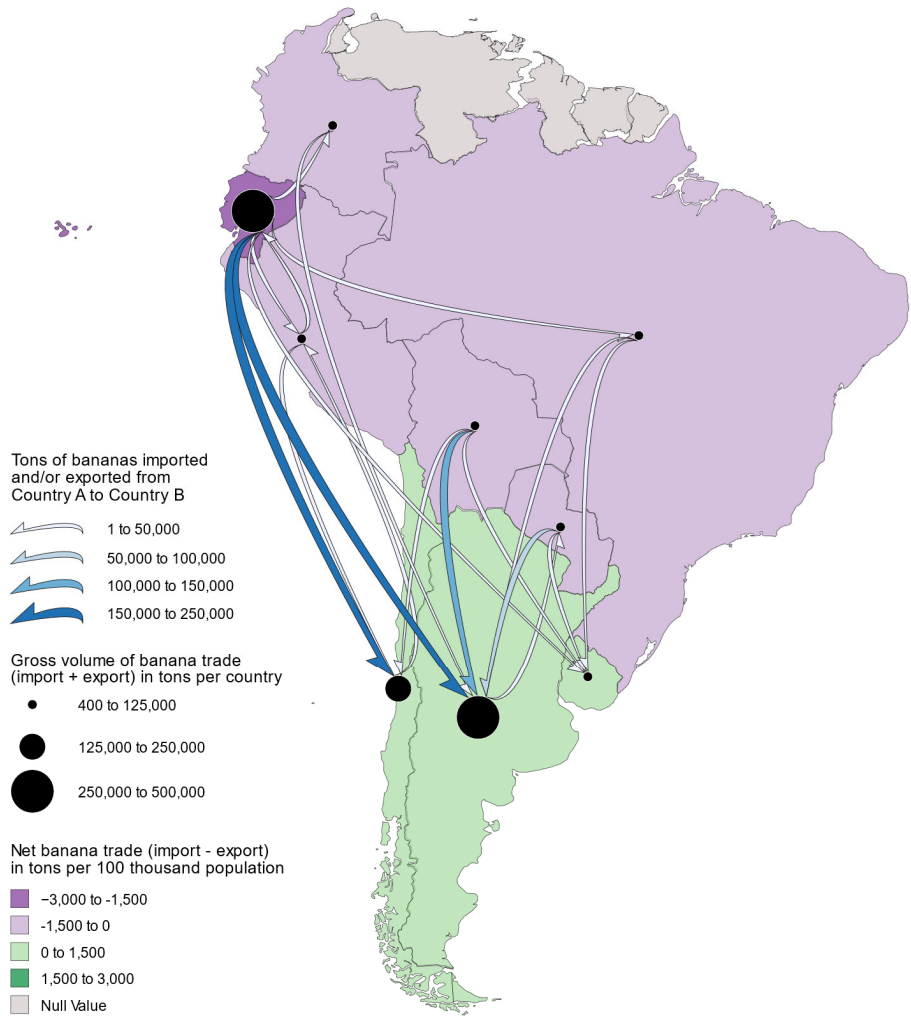

Banana trade between countries in South America

**Figure 3: Banana trade flows between countries in South America in 2019.**

## 4.2.    Scenario 2: Family migration in the United States between 1887 and 1924

Figure 4 illustrates the number of families migrated between states from 1887-1924 derived from a population-scale family tree data set (Koylu et al., 2020). The choropleth map depicts migration efficiency, which represents the net migration of the state (out-migrant families subtracted from in-migrant families) divided by the sum of in and out-migrant families. A divergent color scheme is used to distinguish the net importer states (blue) from the net exporters (red). The graduated node symbols represent the gross volume of migrant families, which is the sum of in- and out-migrant families per state. Finally, the graduated flow symbols represent the number of families migrating between pairs of states. Predominant flows are between adjacent states and from East to West. Particularly, Oklahoma and California appears to have a great inflow of families during this period, most likely because of the oil boom with the first column of oil being discovered in 1897 (Aoghs.org Editors, 2020) and the westward expansion.

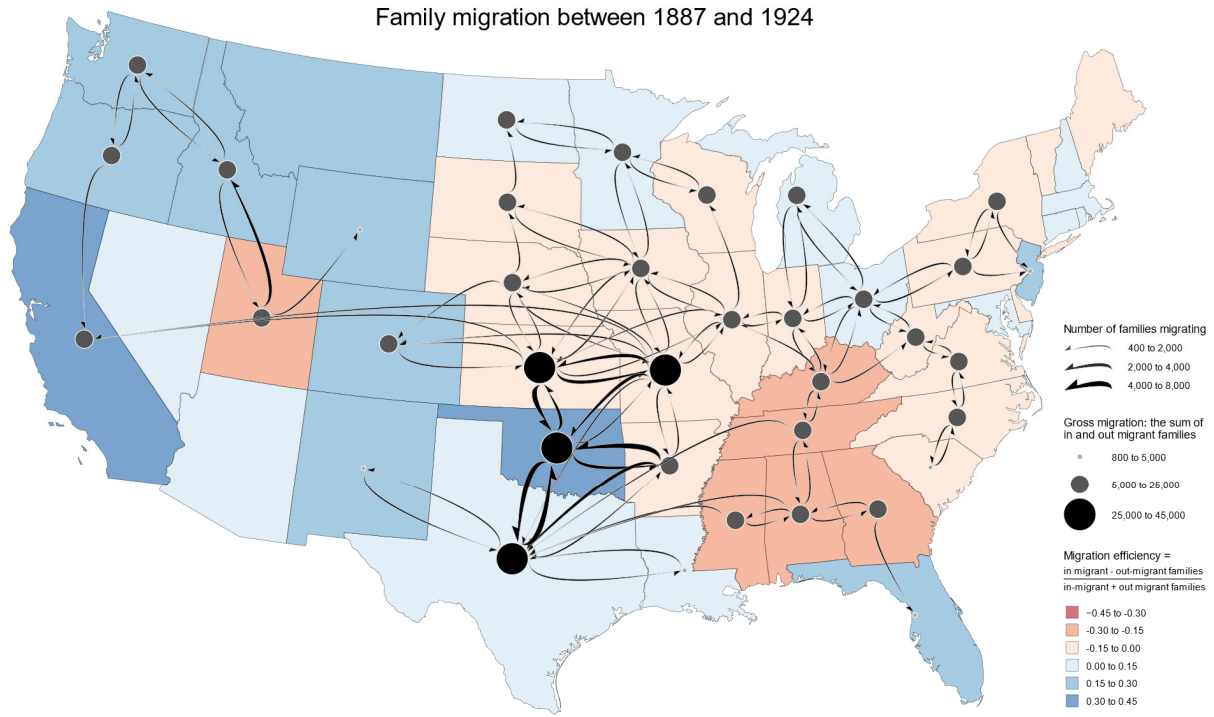

**Figure 4: Migration of families between 1887 and 1924 in the United States.**

## 4.3. Scenario 3: Bike trips in Chicago in April 2021

Figure 5 illustrates bike trips in Chicago in April 2021 (Divvy, 2021). Node symbols are placed at bike stations and classified to represent gross volume of flows (the sum of pick-ups and drop offs) to distinguish how each bike station is used. The Lake Shore Dr. and Monroe St. bike station is selected to highlight flows from and to this station while the rest of the flows are shown transparent. The figure uses a dark basemap with light and neon colors for nodes and flows to provide increased contrast. The selected station is nearby attractions such as Maggie Daley Park, the Art Institute of Chicago, and Cloud Gate. The major in flows to this station come from the Chicago Field Museum, the Shedd Aquarium, and the Alder Planetarium from southern part of downtown. The largest number of bike trips for this station are from and to the Navy Pier bikeshare station which is to the northwest. Other larger out flows from Lake Shore Dr. and Monroe St. station are located along a bike path that goes along Lake Shore Dr. that is to the northwest of this location.

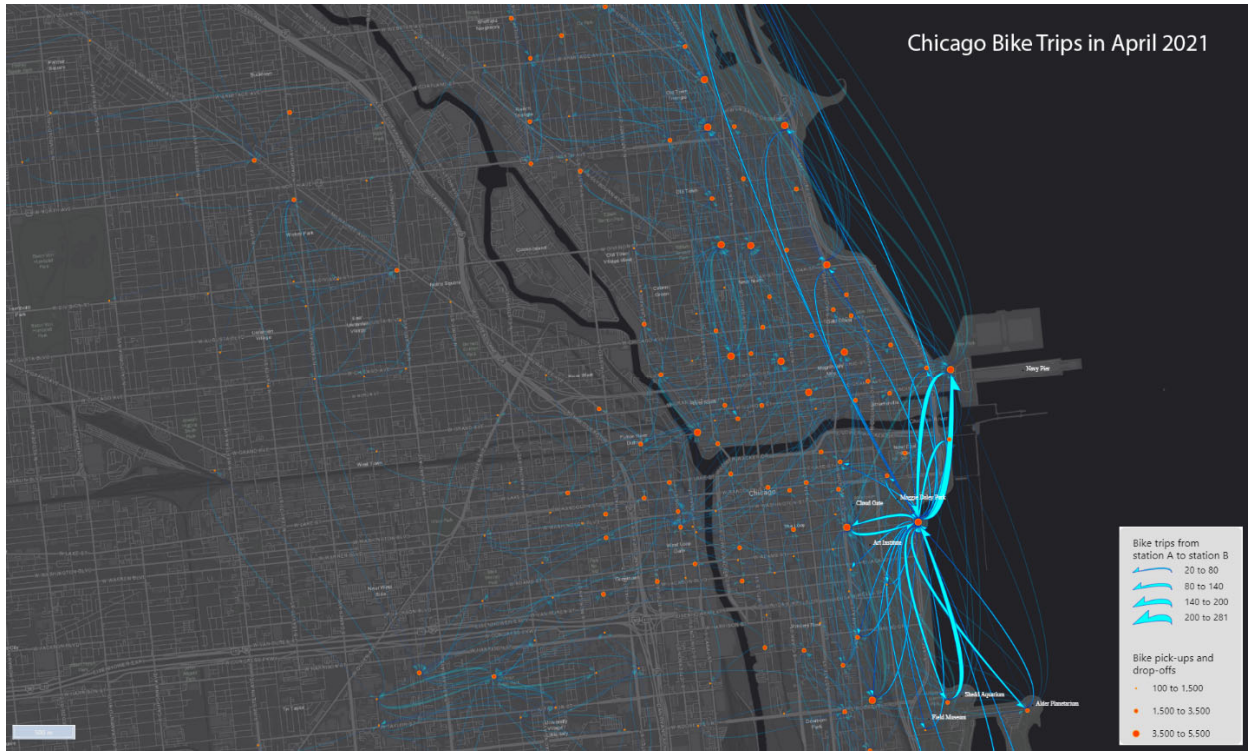

**Figure 5: Selected bike trips in Chicago in April 2021 from and to Lake Shore Dr. and Monroe St. bike station using the curve with half-arrow style.**

## 5.    Conclusions and future work

Our overarching goal for FlowMapper is to provide users greater customizability and design flexibility that could result in an interactive map and a static map product that requires minimal touch-up in other graphics software. Our goal is to address current limitations and make enhancements so that FlowMapper can be used in the professional cartographic workflow to make polished static and interactive maps. In this article, we summarized the main contributions of FlowMapper. First, users can upload and process their own data to produce, customize and share flow maps. The ability to save data, cartographic design and map elements in a project file allows users to easily share their data and/or cartographic design with others. Second, users can customize the flow line symbology by including options to change the flow line style, width, and coloring. We introduce algorithms for drawing curved line styles with varying thickness along a flow line, which reduces the visual cluttering and overlapping by tapering flow lines at origin and destination points. The ability to customize flow symbology supports a range of flow map reading tasks such as comparing flow magnitudes and directions and identifying flow and location clusters that are strongly connected with each other. Third, users can add and customize supplementary layers such as node symbol, choropleth, and base maps to support contextualizing flow patterns with location references and characteristics such as gross flow, net-flow ratio or a locational attribute such as population density and median income. FlowMapper also supports user interactions to zoom, filter, and obtain details-on-demand functions to acquire information on flows, nodes, and regions. Finally, the web-based architecture of FlowMapper supports server-side computational capabilities to process and summarize large flow data to reveal natural patterns of flows.

There are several future directions we plan to pursue and address the limitations of FlowMapper. Asymmetric curvature introduced in this manuscript reduces overlapping at nodes by tapering the flow curves at origins and destination ends. We apply the same arbitrary curvature to all flows, which has been found to increase visual clutter in certain cases (Xu et al., 2012) such as drawing flows from geographically close origins to geographically close destinations. Thus, each algorithm for drawing curved flows may produce visual clutter elsewhere in a flow map layout and may violate different design principles. To give users more flexibility to choose flow line algorithms based on the desired map reading tasks, we plan to implement more flow line designs including origin-destination coloring and partial lines, symmetric curves and curves with varying and interactively controllable symmetry, and the force-directed algorithm (Jenny et al., 2017). We also plan to implement undirected and one-way directed (or net) flow maps as alternative map types. We plan to expand the sharing of flow maps by enabling users to host and process data and share their flow maps on the cloud. We currently support a limited number of map projections. We plan to incorporate more projections and the ability to automatically select an appropriate projection based on the spatial extent of the data. While we currently support flow normalization, we plan to integrate computational capabilities such as multi-scale flow mapping (Zhu et al., 2019), and regionalization (Guo, 2008). FlowMapper does not support the visualization of temporal flow data. We plan to extend FlowMapper to incorporate alternative time-series flow visualizations such as flow map animations and small multiples. We also plan to support the design decisions for FlowMapper by developing a pattern typology for OD flow data and patterns (Dodge, 2019; Dodge, Weibel, & Lautenschutz, 2008; Purchase et al., 2008). A typology will help users to identify types of patterns with recommended symbology options.

## Software

The front end of FlowMapper is built with HTML, CSS, JavaScript and the bootstrap template. OpenLayers is used for the mapping framework including the base maps and basic interactive map functions, and D3 (Data-Driven Documents) (Bostock, Ogievetsky, & Heer, 2011) is used to create the interactive maps that include polygon, point and flow symbol vectors. FlowMapper utilizes a Java back-end design connected with a PostgreSQL/PostGIS database with an Apache Tomcat server. Currently, flow data transformation and normalization features can be employed using the back-end server. Back-end design will enable future data hosting, and flow data processing such as regionalization (Guo, 2008) and multi-scale flow mapping (Zhu et al., 2019).

## Data availability statement

All the source code to produce both back-end and front-end development of FlowMapper and the example flow data sets used in this article are publicly available with a DOI at: https://doi.org/10.6084/m9.figshare.14593380.v3. The data sets used in this study do not contain human subjects, and they are in the public domain. Banana trade flow data between countries in South America from 2019 were derive from Food and Agriculture Organization of United Nations website: http://www.fao.org/faostat/en/#data/TM (FAO, 2021). The country polygons were obtained from Natural Earth Data (Natural Earth, 2021). Historical family migration data between states from 1887-1924 were derived from a population-scale family tree data set produced by Koylu *et al.* (2020). Bike trip data in Chicago throughout the month of April 2021 were derived from https://www.divvybikes.com/system-data (Divvy, 2021). All the source code and materials for FlowMapper are also published on this GitHub link: https://github.com/geo-social/flowmapper.

## Acknowledgements

The authors thank Daniel Stephen, Dr. Brooke Marston, and Dr. Yalong Yang for their valuable feedback and suggestions. The authors also thank the department of Geographical and Sustainability Sciences at the University of Iowa for supporting the development of FlowMapper as a teaching material to be used in geography, GIS, and other relevant courses. The authors thank Mert Erdemir, Beichen Tian and Hoeyun Kwon for their kind support and feedback.

## Disclosure statement

# Figure captions and alt text

**Figure 1:** Flow line styles (a) Curve with half-arrow (b) Straight with half-arrow (c) Tapered (d) Teardrop

**Figure 1 Alt Text:** Four flow symbol design FlowMapper currently supports: Curve with half-arrow, tapered curve, teardrop curve, and straight-line with half-arrow. All the curves are asymmetric to enhance readability of directions in two-way flow maps.

**Figure 2:** Scaling and legend alternatives for flow symbols (a) a single color (black) flow symbol with line thickness scaled proportional to flow volume. (b) color and line thickness scaled proportional to flow volume (c) a classified flow thickness and color with five classes using a quantile classification.

**Figure 2 Alt Text:** Figure 2 illustrates proportional symbol legends with a single color (Figure 2.a) and a continuous (unclassed) color scheme from white to blue with a gray flow stroke (Figure 2.b). Proportional symbol legends include three values: minimum, average, and maximum. Figure 2.c illustrates a classified flow thickness and color with five classes using a quantile classification.

**Figure 3:** Banana trade flows between countries in South America in 2019.

**Figure 3 Alt Text:** Banana trade flows between countries in South America from 2019 (FAO, 2021). We depict the net-flow ratio for each country on the choropleth map, the total flow volume (import + export) on the proportional symbol map, and volume of trade flows between countries using the teardrop style. Countries with a warmer climate that are closer to the equator export more bananas to countries in the south of the continent that have less suitable climate for producing bananas.

**Figure 4:** Historical migration of families between 1887 and 1924 in the United States.

**Figure 4 Alt Text:** The number of families migrated between states from 1887-1924 derived from a population-scale family tree data set (Koylu et al., 2020). Predominant flows are between adjacent states and from East to West. Particularly, Oklahoma and California appears to have a great inflow of families during this period, most likely because of the oil boom with the first column of oil being discovered in 1897 (Aoghs.org Editors, 2020) and the westward expansion.

**Figure 5:** Selected bike trips from and to Lake Shore Dr. and Monroe St. bike station using curve with half-arrow symbol.

**Figure 5 Alt Text:** In this map, the Lake Shore Dr. and Monroe St. bike station is selected to highlight flows from and to this station. Figure 5 uses the curve symbol with half-arrow to reinforce the perception of directions, and a dark basemap with light and neon colors to provide increased contrast. The selected station is nearby attractions such as Maggie Daley Park, the Art Institute of Chicago, and Cloud Gate. The major in flows to this station come from the Chicago Field Museum, the Shedd Aquarium, and the Alder Planetarium from southern part of downtown. The largest number of bike trips for this station are from and to the Navy Pier bikeshare station which is to the northwest. Other larger out flows from Lake Shore Dr. and Monroe St. station are located along a bike path that goes along Lake Shore Dr. that is to the northwest of this location.

# Appendices

## 1. Flow path symbology and algorithms

Here we illustrate the asymmetric Bezier curve with a half-arrow connecting an origin node to a destination node (Figure A.1). The flow symbol is curvy at the origin and straight at the destination end to increase readability of flow maps by tapering the flow lines that are converging to and diverging from nodes. Curved flow line with half-arrow symbology consists of two cubic Bezier curves between the origin and destination points with an additional half-arrow drawn at the destination. We paint the area between the two curves and the half-arrow to generate the flow line style. We calculate the coordinate points explained below using the algorithm described in Appendix Table A.1. There are four major steps in curve with half-arrow algorithm:

1- We first identify the origin and destination points of the flow symbol on the periphery of the node symbols (circles) instead of directly drawing the flow symbol between the coordinates (centroids) of the origin-destination points. We name these points as P0 (origin) and P3 (destination), which are at the intersection of the periphery of the node symbol (circle) and the imaginary straight line between P0 and P3.

2- We then draw a straight line from P0 to an offset point (P1) which is perpendicular to the imaginary straight line between P0 and P3. Because curves converge at the destination end touching the circle of the node symbol, the offset point allows the flows to be drawn away from the origin node and reduce cluttering of potential overlap between the flows that leave the node and flows that arrive at the node.

3- We draw the first Bezier cubic curve that will produce the outer edge of the flow symbol. A cubic Bezier curve is drawn with three points. While the first two points are control points, the last point is the end point for the curve. To draw a half-arrow, we estimate the end point (P2) for the outer curve to be before the destination point (P3) on an imaginary Bezier curve line to the destination point (P3). To draw the asymmetric curve, we now have the start point (P1), the two control points: CP1_C1 (control point 1 for curve 1) and CP2_C1 (control point 2 for curve 1) that are marked with a red stroke color, and the end point (P2).

4- We draw the half-arrow using two straight lines. The first line is the straight line between P2 and EP3 (elbow point for the arrow), and the second line is the destination point (P3). Now, the outer edge of the flow symbol is drawn.

5- Drawing of the inner flow symbol is straightforward since it does not include an offset or a half-arrow. The inner flow is also a cubic Bezier curve that starts from P3, uses two control points of CP1_C2 and CP2_C2, and the end point P0.

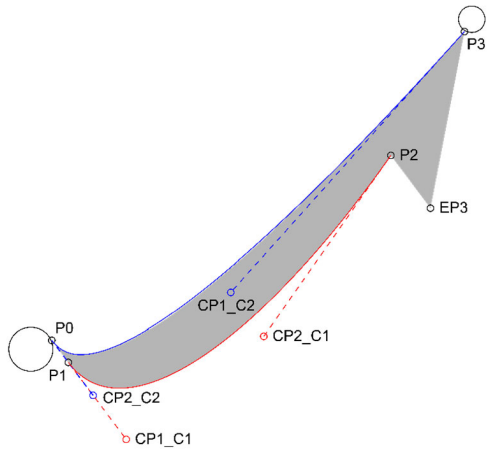

**Figure A.1: Bezier curve with half-arrow. P0 and P3 are origin and destination points. CP points are control points for the inner (blue) and outer (red) Bezier curves. P1 and P2 are offset points for origin and destination, respectively. EP3 is the elbow point for drawing the half-arrow.**

In tapered flow line symbol, the only difference is in the drawing of the outer Bezier curve. We simply remove the destination offset point (P2) and the elbow point (EP) and draw a cubic curve from P1 to P3 using the same control points P1_CP1 and new control point calculated directly from the destination point P3. Teardrop flow symbol is the reverse of the tapered symbol. In teardrop design, the origin and destination points are flipped, and the origin offset point P1 (which now would be a destination offset point) is removed. Teardrop has been found to be effective for direction tasks in flow map reading (Koylu & Guo, 2017; Ware et al., 2014). Straight flow line symbol includes a destination offset and an elbow point to draw the half-arrow. Straight flow with half-arrow has been used extensively in flow mapping (Ilya Boyandin, 2021; Guo, 2009). Below is the JavaScript code for producing half-arrow Bezier curve flow line symbolization.

**Table A.1: The algorithm to construct Bezier curve with half-arrow.**

| |
| --- |
| **Input:** x0, y0: x and y coordinates of the origin location |
| x3, y3: x and y coordinates of the destination location |
| flowSize: Flow thickness (in pixel size) determined by the flow magnitude (e.g., volume) |
| sourceRadius: Size of origin symbol (circle) |
| targetRadius: Size of destination symbol (circle) |
| righthandrule: if true then draw flows using the right-hand traffic rule, if false  the draw using the left-hand traffic rule |
| **Output:** Curved path with half-arrow |
| Corresponding points calculated for Figure 2: |
| P1: xc1, yc1 |
| P2: xc2, yc2 |
| CP1_C1:xc13rd, yc13rd |
| CP2_C1: xc23rd, yc23rd |
| EP3: x3elbow, y3elbow |
| CP1_C2: x23rd, y23rd |
| CP2_C2: x13rd, y13rd |

```
1    function drawCurve (x0, y0, x3, y3, flowSize, sourceRadius, targetRadius, righthandrule){
2
3            var arrowlen = 2.42;
3            var arrowwidthconstant = flowSize * 1.1;
4
5            // adjusting half-arrow size
6            if(flowSize < 10 && flowSize >= 8){
7                    arrowlen = 2.64;
8                    arrowwidthconstant = flowSize * 1.2;
9            }
10           else if(flowSize < 8 && flowSize >= 6){
11                   arrowlen = 3.08;
11                   arrowwidthconstant = flowSize * 1.4;
12           }
13           else if(flowSize < 6 && flowSize >= 4){
14                   arrowlen = 4.4;
15                   arrowwidthconstant = flowSize * 2;
16           }
17           else if(flowSize < 4 && flowSize >= 3){
17                   arrowlen = 6.6;
18                   arrowwidthconstant = flowSize * 3;
19           }
20           else if(flowSize < 3 && flowSize >= 2){
21                   arrowlen = 8.8;
22                   arrowwidthconstant = flowSize * 4;
23           }
23           else if(flowSize < 2){
24                   arrowlen = 11;
25                   arrowwidthconstant = flowSize * 5;
26           }
27
28           var ndsize0 = sourceRadius;
29           var ndsize3 = targetRadius;
30           var dx = Math.abs(x3 - x0);
30           var dy = Math.abs(y3 - y0);
31
32           var len = Math.sqrt(dx * dx + dy * dy);
33           if (len < (ndsize0 + ndsize3) * 1.2)
34                   return;
35
36           // Shorten the length so that the line only touches the node circle
36           var haselbow = true;
37           if (haselbow) {
38                   x0 = x0 + (x3 - x0) * ndsize0 / len;
39                   y0 = y0 + (y3 - y0) * ndsize0 / len;
40                   x3 = x3 - (x3 - x0) * ndsize3 / (len - ndsize0);
41                   y3 = y3 - (y3 - y0) * ndsize3 / (len - ndsize0);
42           }
43
44           // Four corners of a flow line
44           var xc1 = null, yc1 = null, xc2 = null, yc2 = null;
45           // Four points to round the head
46           var x3elbow1 = null, y3elbow1 = null;
46           var sign = -1;
```

```
47          var xdelta = null, ydelta = null;
48          xarrowdelta = null, yarrowdelta = null, xgap = null, ygap = null;
49          var gap = flowSize * 0.05;
50          if (y0 == y3) { // horizontal
51              xdelta = 0;
52              ydelta = flowSize / 2;
53              xarrowdelta = 0;
54              yarrowdelta = arrowwidthconstant / 1.0;
55              xgap = 0;
55              ygap = gap;
56          } else if (x0 == x3) { // vertical
57              ydelta = 0;
58              xdelta = flowSize / 2;
59              yarrowdelta = 0;
60              xarrowdelta = arrowwidthconstant / 1.0;
60              xgap = gap;
60              ygap = 0;
61          } else {
62              var v = (x3 - x0) / (y0 - y3);
63              xdelta = flowSize / 2.0 /  Math.sqrt(1 + v * v);
64              ydelta = Math.abs(xdelta * v);
65              xarrowdelta = arrowwidthconstant / Math.sqrt(1 + v * v);
66              yarrowdelta = Math.abs(xarrowdelta * v);
66              xgap = gap / Math.sqrt(1 + v * v);
67              ygap = Math.abs(xgap * v);
68              if (v < 0)
69                      sign = 1;
70          }
70          x0 = (y0 > y3) ? x0 + xgap : x0 - xgap;
71          x3 = (y0 > y3) ? x3 + xgap : x3 - xgap;
72          y0 = (x0 > x3) ? y0 - ygap : y0 + ygap;
73          y3 = (x0 > x3) ? y3 - ygap : y3 + ygap;
74
75          if(righthandrule)
76          {
76              xc1 = (y0 > y3) ? x0 + xdelta / 2 : x0 - xdelta / 2;
77              yc1 = (x0 > x3) ? y0 - ydelta / 2 : y0 + ydelta / 2;
78          }
79          yc2 = (x0 > x3) ? y3 - ydelta + arrowlen * xdelta * sign : y3 + ydelta -
80 arrowlen * xdelta * sign;
81          xc2 = (y0 > y3) ? x3 + xdelta + arrowlen * ydelta * sign : x3 - xdelta -
82 arrowlen * ydelta * sign;
83          x3elbow1 = (y0 > y3) ? x3 + xarrowdelta + arrowlen * ydelta * sign : x3
84 - xarrowdelta - arrowlen * ydelta * sign;
85          y3elbow1 = (x0 > x3) ? y3 - yarrowdelta + arrowlen * xdelta * sign : y3
86 + yarrowdelta - arrowlen * xdelta * sign;
87
88          var arcxdelta = xdelta * len / 4 / flowSize;
89          var arcydelta = ydelta * len / 4 / flowSize;
90          var x13rd = (y0 > y3) ? x0 + arcxdelta : x0 - arcxdelta;
90          var y13rd = (x0 > x3) ? y0 - arcydelta : y0 + arcydelta;
91          var x23rd = (y0 > y3) ? x0 + (x3 - x0) / 3 + arcxdelta :
92 x0 + (x3 - x0) / 3 - arcxdelta;
93          var y23rd = (x0 > x3) ? y0 + (y3 - y0) / 3 - arcydelta :
```

```
y0 + (y3 - y0) / 3 + arcydelta;
        arcxdelta = arcxdelta + xdelta;
        arcydelta = arcydelta + ydelta;
        var xc13rd = (y0 > y3) ? x0 + arcxdelta : x0 - arcxdelta;
        var yc13rd = (x0 > x3) ? y0 - arcydelta : y0 + arcydelta;
        var xc23rd = (y0 > y3) ? x0 + (x3 - x0) / 3 + arcxdelta : x0 + (x3 - x0)
/ 3 - arcxdelta;
        var yc23rd = (x0 > x3) ? y0 + (y3 - y0) / 3 - arcydelta : y0 + (y3 - y0)
/ 3 + arcydelta;

        }else
        {
        //left-hand traffic rule
        xc1 = (y0 < y3) ? x0 + xdelta / 2 : x0 - xdelta / 2;
        yc1 = (x0 < x3) ? y0 - ydelta / 2 : y0 + ydelta / 2;
        yc2 = (x0 < x3) ? y3 - ydelta + 2.5 * xdelta * sign :
y3 + ydelta - 2.5 * xdelta * sign;
        xc2 = (y0 < y3) ? x3 + xdelta + 2.5 * ydelta * sign :
x3 - xdelta - 2.5 * ydelta * sign;

        x3elbow1 = (y0 < y3) ? x3 + xarrowdelta + 2.5 * ydelta * sign :
x3 - xarrowdelta - 2.5 * ydelta * sign;
        y3elbow1 = (x0 < x3) ? y3 - yarrowdelta + 2.5 * xdelta * sign :
y3 + yarrowdelta - 2.5 * xdelta * sign;

        var arcxdelta = xdelta * len / 4 / flowSize;
        var arcydelta = ydelta * len / 4 / flowSize;
        var x13rd = (y0 < y3) ? x0 + arcxdelta : x0 - arcxdelta;
        var y13rd = (x0 < x3) ? y0 - arcydelta : y0 + arcydelta;
        var x23rd = (y0 < y3) ? x0 + (x3 - x0) / 3 + arcxdelta :
x0 + (x3 - x0) / 3 - arcxdelta;
        var y23rd = (x0 < x3) ? y0 + (y3 - y0) / 3 - arcydelta :
y0 + (y3 - y0) / 3 + arcydelta;
        arcxdelta = arcxdelta + xdelta;
        arcydelta = arcydelta + ydelta;

        var xc13rd = (y0 < y3) ? x0 + arcxdelta : x0 - arcxdelta;
        var yc13rd = (x0 < x3) ? y0 - arcydelta : y0 + arcydelta;
        var xc23rd = (y0 < y3) ? x0 + (x3 - x0) / 3 + arcxdelta : x0 + (x3 - x0)
/ 3 - arcxdelta;
        var yc23rd = (x0 < x3) ? y0 + (y3 - y0) / 3 - arcydelta : y0 + (y3 - y0)
/ 3 + arcydelta;
        }

        return "M" + x0 + "," + y0
+ " L" + xc1 + "," + yc1
+ " C" + xc13rd + ","+ yc13rd + " " +xc23rd+ "," +yc23rd+ " "+ xc2 + "," + yc2
+ " L" + x3elbow1 + "," + y3elbow1
+ " L" + x3 + "," + y3
+ " C" + x23rd + "," +y23rd+ " " +x13rd+ "," +y13rd+ " " + x0 + "," + y0;
}
```

## 2. Flow normalization

Flow normalization consists of two steps. We first calculate an expected flow matrix based on a statistical null model. Second, we subtract the flow expectation based on the null model from the observed volume for each of the flows.

### 2.1. Calculating flow volume expectation using a null model

There are numerous functions that can be used to calculate flows between locations (nodes) in a spatial network. The most common parameters to consider in these models are gross flow volume per location, population, and geographic proximity. In this section we present two null models: adjusted flow volume and double-constrained gravity model.

#### 2.1.1 Adjusted flow volume model

Adjusted flow volume model requires only flow data to calculate the expected values for each flow. This feature is available in FlowMapper.org version 1.0.5. We use an adjusted flow volume formula to calculate the expected number of flows between locations.

$$\text{Expected Flow (O, D)} = F_O \, F_D \, f\,(O, D) \,/\, (F_S{}^2 - \sum_{i=0}^{n} F_i{}^2 \,)$$

where EF (O, D) is the expected number of flows between origin O and destination D, $F_O$ is the observed number of flows between origin O and its connections, $F_D$ is the observed number of flows between destination D and its connections, f (O, D) is the observed number of flows between origin O and destination D, $F_S$ is the observed number of flows between all locations, and $\sum_{i=0}^{n} F_i{}^2$ is used to remove within-location expectations if there are any.

#### 2.1.2 Double-constrained gravity model

Alternative to flow adjusted expectation model, one can use a double-constrained gravity model to consider the effect of geographic proximity and population on generating flows. Gravity model requires the node file (or the geojson shapes of the regions) and the flow data to calculate the expected values for each flow. FlowMapper 1.0.5 currently does not support the gravity model, however, this feature will be available in future revisions. The gravity model enforces the sum of expected flows from an origin is equal to the observed and the sum of expected flows to a destination is equal to the observed volume of flows to that destination (Roy & Thill, 2004).

$$E_{ij} = A_i * O_i * B_j * D_j * D_{ij}{}^{-beta}$$

where $O_i$ and $D_i$ are the sum of out-flows and in-flows of location i, $A_i$ and $B_i$ are the balance factors that are calculated by the following iteration. The distance decay function is square and uniform for all locations, each node (state) has a different set of parameters.

$$A_i = 1/ \sum_{i=0}^{n} \sum_{j=0}^{n} (B_j D_j \ * \ D_{ij}{}^{beta})$$

$$B_i = 1/\sum_{i=0}^{n}\sum_{j=0}^{n}\left(A_j O_j \ * \ D_{ij}{}^{beta}\right)$$

While positive modularity values indicate observed flows are above the expectation, negative values indicate flows are less than the expectation.

## 2.2.    Calculation of modularity

Finally, we transform raw volume of flows into modularity flows using the following formula.

MOD (O, D) = Observed Flow − Expected Flow

where OF is actual (observed) number of flows, and EP is expected number of flows on link O-D. Using this formula, the raw counts of flows are transformed into a modularity graph, in which the weight of a link represents the modularity between two locations. If modularity value is positive the flow is above expectation, if the value is negative the flow is below expectation.

## 2.3.    Case study: family migration between 1887 and 1924 with the gravity model

Figure A.2 illustrates modularity transformed family migration between 1887 and 1924 using the double-constrained gravity model. When both the effect of geographic distance and the differences of gross flow volumes between states are considered, the flow map reveals spatial trends that highlight strong two-way connections between Utah and Idaho, Arkansas and Oklahoma, Missouri and Kansas, and Iowa and Nebraska. These connections show how families moved back on forth between these adjacent states, which may reflect the processes of family settlement, child and elderly care, and return migration. Also, flows to the west (i.e., Washington, Oregon, and California) and to Texas from Mid-West, the southeast become more prevalent.

Modularity of migration between 1887 and 1924 that represents migration of families that are above expectation of the gravity model

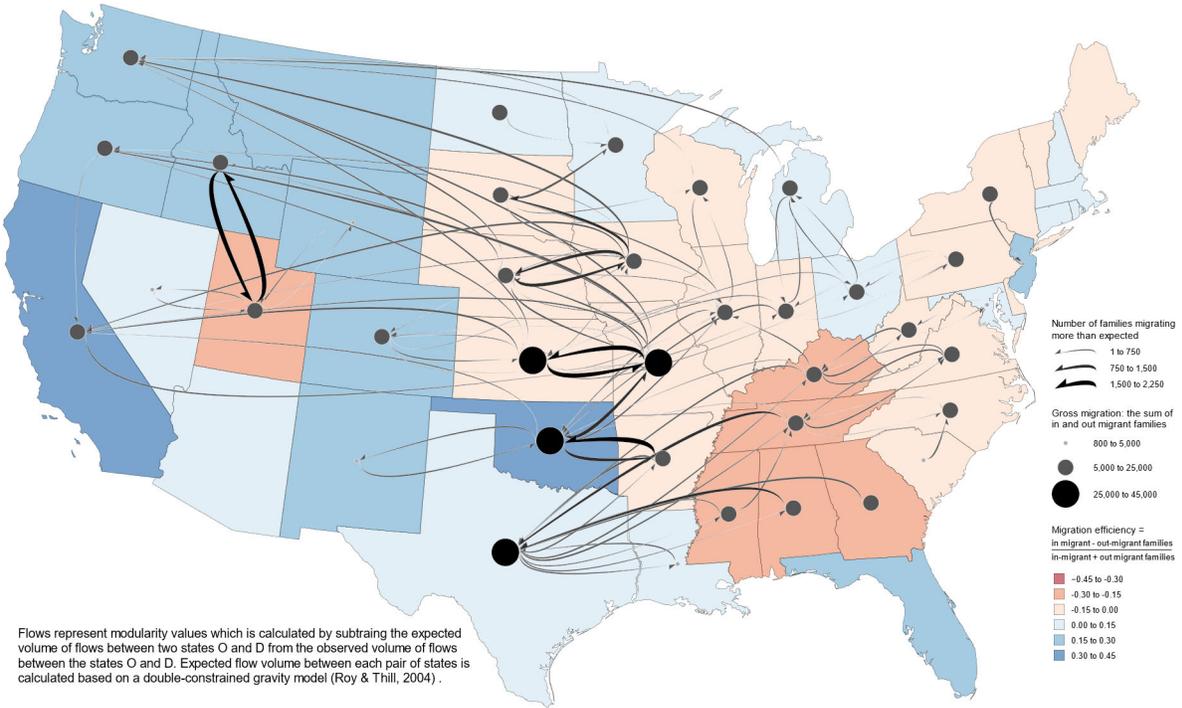

**Figure A.2: Migration of families between 1887 and 1924 that are more than expected based on the gravity model.**

## 3. User interface and experience

We describe the user interface (UI) and the user experience (UX) of FlowMapper from a general user's perspective. Upon accessing FlowMapper, the user will see a base map in the center of the application, a navigation bar at the top, and a flow map settings panel on the left of the interface that is collapsible to provide more space for the map interface (Figure A.3.a). The map is where the flow dataset will be visualized after it is uploaded, and the flow map settings are given. Using the map, the user can pan around, zoom in on areas, and zoom out. To change the base map, the user will use the map settings panel under the 'Base Map' tab and select the desired base map from the Base Map drop-down menu (Figure A.3.b.1). Within the 'Base Map' tab, the user can also change the base map's opacity (Figure A.3.b.2) the map's projection using the drop-down 'Projection' menu (Figure A.3.b.3). The user can also add a title (Figure A.3.b.4), a north arrow (Figure A.3.b.5), a projection label (Figure A.3.b.6), and custom references (Figure A.3.b.7) to the map.

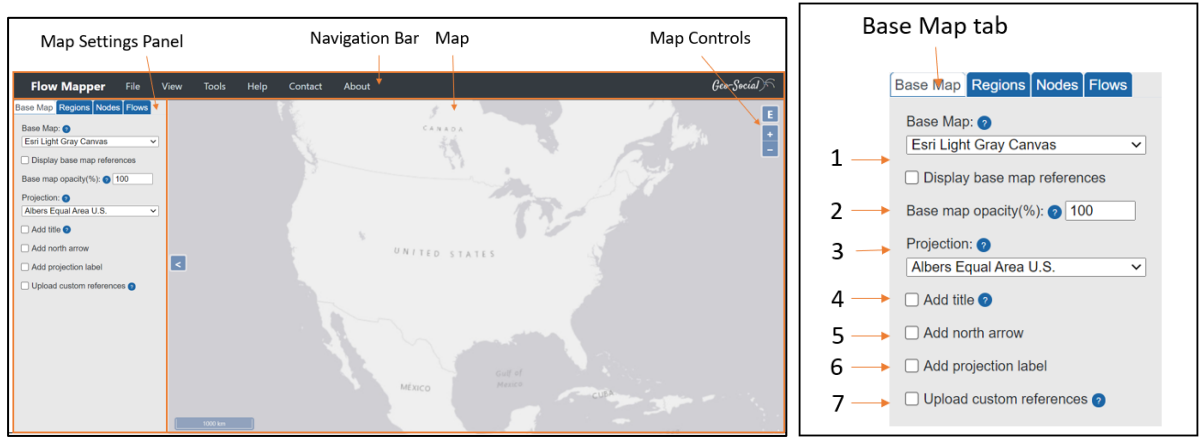

**Figure A.3: (a) FlowMapper user interface and user experience and (b) base map controls**

To start uploading flow data, the user can use regions (polygons), nodes (points), and flows (flow lines) to visualize their dataset. While regions are optional, nodes and flows are required for the flow map application. The regions are uploaded through the 'Regions' tab of the map settings panel. The regions dataset must be in a geojson format and the polygons within it should have an ID that joins to a csv dataset to map any relevant attribute data. Figure A.4.a shows an example dataset of a geojson uploaded for the polygon dataset and a csv dataset uploaded for the attribute dataset. Below the uploaded datasets are where the joins occur. Here the user would join the geojson ID to the csv ID. Once they are joined, the user can select a field to display the attribute data and then choose symbology options below in the panel. The symbology options are the data classification method, the color scheme, the number of classes for the classification (quantiles, equal intervals, natural breaks, linear scale, and manual breaks), fil-opacity (%), stroke color, stroke width. Once these options are filled out, the legend options can be formatted, such as the data layer title and the number of decimal places that should be displayed within the legend. Lastly, to add the dataset to the map with the symbology and legend options set, the user would select the lower left button "Map Regions".

Flow Mapper requires users to upload a node (point) file to determine the origin and destination point coordinates for flows. The nodes data set should be uploaded from within the 'Nodes' tab as a csv. One can create point (node) data using Polygons to Points tool under "Tools" menu, which generates a point data set from the centroid of a polygon data set (regions) in JSON format. The data set should include a node ID which joins on a one-to-many relationship to the origin ID and destination ID of the flows csv dataset. The nodes dataset must also contain an "X" coordinate field and a "Y" coordinate field in latitude and longitude. The flow volume can be mapped and then symbolized with the same options that the regions tab provides. Like within the Regions tab, the legend can be formatted by title and number of decimal places. To create to the nodes on the map, the user would select "Map Nodes" at the bottom left of the nodes tab within the panel (Figure A.4.b).

To map the flows, the user should select the 'Flows' tab within the Map Settings panel. The flow dataset should be uploaded as a csv, and then the user can input the fields that should be joined to the nodes, the flow origin ID and the flow destination ID. Next, the user would input the flow volume field and select symbology options. The symbology options are similar to those discussed from within the 'Regions' tab, except that the flows can also be visualized to show the top *n* number of flows for filtering purposes and the style of flow can be chosen. The style of flow can either curved half arrow, straight half arrow, tapered, or teardrop (Figure 1). Lastly, the user would input the legend formatting options and then select 'Map Flows' at the bottom left of the 'Flows' tab (Figure A.4.c).

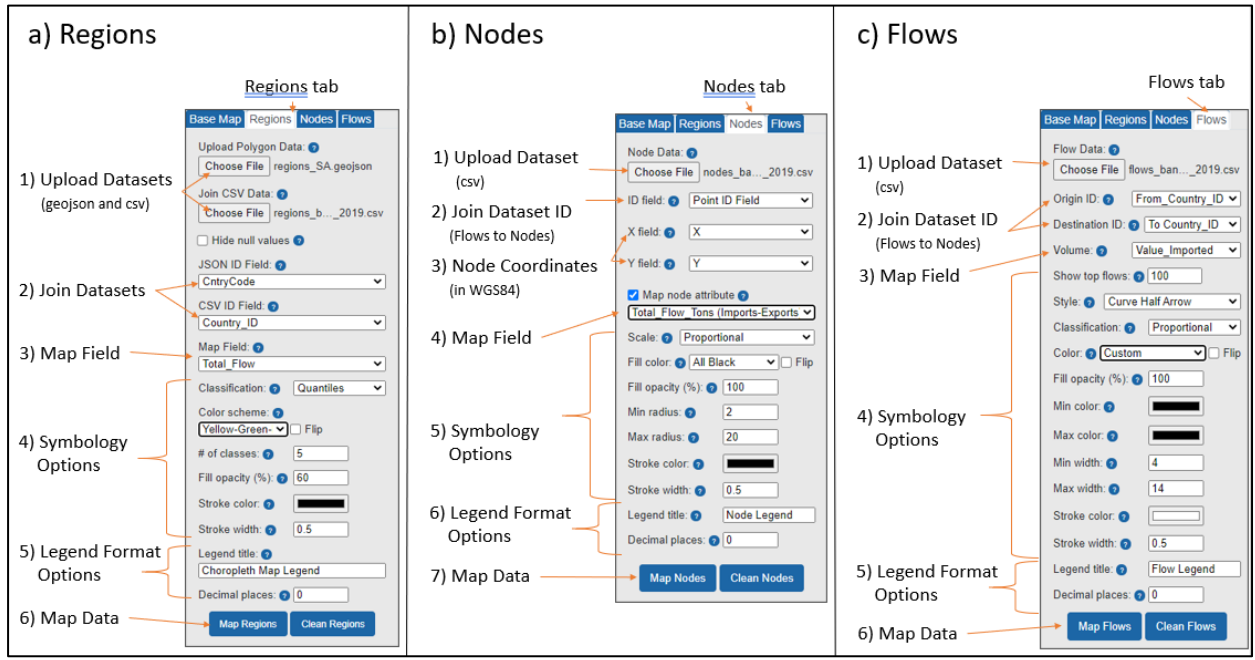

**Figure A.4: Map Settings panel interface showing the Regions tab, the Nodes tab, and the Flows tab.**

If the user wishes to save their current project, they should use the 'File' menu → 'Project' → 'Save Project'. The project will be presented as a json file which will include the joins, the classification or scaling, the symbology, legend settings, and geographic data. The project file will be downloaded locally. If the user wants to open their project in FlowMapper, the user should upload the json file by selecting 'File' menu → 'Project' → 'Load Project'. To export the map as an image, the user can go to the 'File' menu → 'Export' → and select 'Export SVG' to save it as an SVG, or 'Export PNG' to save it as a PNG.

In the navigation menu, the 'About' section provides help and guidance about the user interface and usability of FlowMapper. 'Project Help' gives the user general tips for using the interface, 'Developer Info' shows the Geo-Social Lab's contact information for making suggestions or receiving help on projects. The 'Tutorials' section within the 'About' menu provides the user with video tutorials about using FlowMapper, hoping to resolve users' questions and guide them through using the interface so they can more easily make their own flow map.

## 4. Data format

The datasets that can be loaded into FlowMapper must be in a certain format. First, the polygon dataset for the Choropleth map must be in a geojson or topojson format. For the polygon attribute data, a csv can be joined to the geojson within FlowMapper if both datasets have a common ID to join on. Second, the nodes dataset must be a csv file with X and Y coordinates in WGS84 datum. It must also have a common ID field so the flows can join to the nodes once the flow dataset is also uploaded. The nodes dataset should also have a volume field within it for the node symbology to be displayed within the map. Lastly, FlowMapper requires a flow dataset that should be in csv format. It should have an origin ID and a destination ID. Both ID's should match the common ID in the node csv file. Next, to represent the flow volume, the flow csv file should include a volume field which can represent the total number of flows from and to certain areas. Using an example dataset, Figure A.5 displays the geojson for the polygon features. Outlined in red is the ID, or Country Code, that the attribute data within the csv (Figure A.6) will be joined on. Figure A.7 displays the node csv format with the X, Y, Country_ID, and Total_Volume fields. Figure A.8 shows the flow dataset where the "To Country Code" field and the "From Country Code" field would join to the node dataset to create the flows. The "Value" field represents the quantity of bananas being import or exported, which is equal to the flow volume filed needed for visualization. Once the dataset files are uploaded to FlowMapper, the user can save a project as one json file. If the user closes out of FlowMapper or the page needs to be refreshed, the user can always upload the json project file that will include the datasets settings and flow map visualization settings.

**Figure A.5: Example polygon geojson file with an id attribute to join with a csv file for creating a choropleth map.**

| Country_ID | Country | Item Code | Item | Year | Unit | Imports | Exports | Net_Flow | Total_Flow | Net_Flow_Ratio |
|---|---|---|---|---|---|---|---|---|---|---|
| 19 | Bolivia | 486 | Bananas | 2019 | tonnes | 0 | 109258 | -109258 | 109258 | -1 |
| 58 | Ecuador | 486 | Bananas | 2019 | tonnes | 0 | 476001 | -476001 | 476001 | -1 |
| 21 | Brazil | 486 | Bananas | 2019 | tonnes | 17 | 57004 | -56987 | 57021 | -0.999403728 |
| 169 | Paraguay | 486 | Bananas | 2019 | tonnes | 64 | 66730 | -66666 | 66794 | -0.99808366 |
| 170 | Peru | 486 | Bananas | 2019 | tonnes | 3 | 449 | -446 | 452 | -0.986725664 |
| 44 | Colombia | 486 | Bananas | 2019 | tonnes | 1860 | 2729 | -869 | 4589 | -0.189365875 |
| 9 | Argentina | 486 | Bananas | 2019 | tonnes | 433273 | 0 | 433273 | 433273 | 1 |
| 40 | Chile | 486 | Bananas | 2019 | tonnes | 245634 | 1 | 245633 | 245635 | 0.999991858 |
| 234 | Uruguay | 486 | Bananas | 2019 | tonnes | 50857 | 0 | 50857 | 50857 | 1 |

**Figure A.6: A snapshot of a polygon csv file that contain attributes to create a choropleth map.**

| Country | X | Y | Country_I | Total_Flow_Tons (Imports-Exports) |
|---|---|---|---|---|
| Uruguay | -56.0181 | -32.7995 | 234 | 50857 |
| Peru | -71.3824 | -10.5828 | 170 | -446 |
| Paraguay | -58.4001 | -23.2282 | 169 | -66666 |
| Ecuador | -78.752 | -1.42382 | 58 | -476001 |
| Colombia | -73.0811 | 3.913834 | 44 | -869 |
| Chile | -71.3826 | -37.7307 | 40 | 245633 |
| Brazil | -53.0978 | -10.7878 | 21 | -56987 |
| Bolivia | -64.6854 | -16.7081 | 19 | -109258 |
| Argentina | -65.1798 | -35.3813 | 9 | 433273 |

**Figure A.7: A node csv file that contains the latitude and longitude coordinates of origin and destination locations, and the id field to match with the flow csv file.**

| To Country_ID | To_Country | From_Country_ID | From_Country | Item | Year | Unit | Value |
|---|---|---|---|---|---|---|---|
| 9 | Argentina | 19 | Bolivia (Plurinati | Bananas | 2019 | tonnes | 104428 |
| 9 | Argentina | 21 | Brazil | Bananas | 2019 | tonnes | 26624 |
| 9 | Argentina | 44 | Colombia | Bananas | 2019 | tonnes | 2710 |
| 9 | Argentina | 58 | Ecuador | Bananas | 2019 | tonnes | 232811 |
| 9 | Argentina | 169 | Paraguay | Bananas | 2019 | tonnes | 66700 |
| 21 | Brazil | 58 | Ecuador | Bananas | 2019 | tonnes | 17 |
| 40 | Chile | 9 | Argentina | Bananas | 2019 | tonnes | 0 |
| 40 | Chile | 19 | Bolivia (Plurinati | Bananas | 2019 | tonnes | 2466 |
| 40 | Chile | 44 | Colombia | Bananas | 2019 | tonnes | 0 |
| 40 | Chile | 58 | Ecuador | Bananas | 2019 | tonnes | 242800 |
| 40 | Chile | 170 | Peru | Bananas | 2019 | tonnes | 368 |

**Figure A.8: A flow csv file that contains origin, destination, and flow volume fields.**